\documentclass[12pt]{article} 
\def \pl {\partial}
\def \lf {\left (}
\def \rt {\right )}
\def \gm {\gamma}
\def \Gm {\Gamma}
\def \w {\omega}

\def \comma {\; , \; \;}
\def \bra {\langle}
\def \ket {\rangle}
\def \wbar {\bar{\w}}
\def \in {|{\rm in} \ket}
\def \out {|{\rm out} \ket}

\begin{document}

\begin{titlepage}

\begin{center}

{\hbox to \hsize{\hfill CU-TP-1101}}
{\hbox to \hsize{\hfill RUNHETC-2003-31}}

\bigskip

\vspace{3\baselineskip}

{\Large \bf A First-Quantized Formalism for Cosmological Particle Production}

\bigskip
\bigskip

{\large \sc Alex~Hamilton,\footnote{\tt hamilton@phys.columbia.edu}${}^a$
Daniel~Kabat,\footnote{\tt kabat@phys.columbia.edu}$^{ab}$ {\rm and}
Maulik~Parikh\footnote{\tt mkp@phys.columbia.edu}$^{a}$}

\bigskip
\medskip

${}^a${\it Department of Physics}\\
{\it Columbia University}\\
{\it New York, NY 10027}\\

\bigskip

${}^b${\it Department of Physics and Astronomy}\\
{\it Rutgers University}\\
{\it Piscataway, NJ 08855}\\

\vspace*{1.0cm}
\end{center}

\noindent
We show that the amount of particle production in an arbitrary
cosmological background can be determined using only the late-time
positive-frequency modes. We don't refer to modes at early times, so
there is no need for a Bogolubov transformation.  We also show that
particle production can be extracted from the Feynman propagator in an
auxiliary spacetime.  This provides a first-quantized formalism for
computing particle production which, unlike conventional Bogolubov
transformations, may be amenable to a string-theoretic generalization.

\end{titlepage}

\newpage

\setcounter{page}{2} \setcounter{footnote}{0}

\tableofcontents
\medskip
\parskip 8 pt

\section{Introduction}

The standard approach to cosmological particle production
\cite{birrelldavies,jacobson} involves finding two sets of solutions
to the wave equation, with pair production given by a Bogolubov
transformation between the two sets of modes.  Although this method
has a long and proven record of success, there are reasons to search
for alternative techniques. In particular, the second-quantized
formalism underlying the Bogolubov transformation does not carry over
to string theory, whose standard perturbative formulation is
first-quantized \cite{samir,mathur}. In order to estimate, say, the
spectrum of density fluctuations coming from stringy effects
\cite{brandenberger,niemeyer,kempf,eastheretal1,kaloperetal,eastheretal2},
one would like to have a first-quantized technique in which one can
systematically compute $\alpha'$ corrections \cite{stringprod}.

In this paper we present a straightforward method for computing
pair-production in an arbitrary cosmological background without using
Bogolubov transformations. We develop the method in two steps.  First
we show how to extract pair production from the Feynman propagator
using only the late-time positive-frequency modes. Then we show that
particle production can be derived from the Feynman propagator of an
auxiliary spacetime, without knowledge of the late-time modes.  Thus
we can compute particle production in a completely first-quantized
framework, without any reference to mode solutions.  Similar
first-quantized techniques have been successful in calculating pair
creation from horizons \cite{tunnel,newcoords}, where again Bogolubov
transformations are usually used.

This paper is organized as follows. In section 2 we briefly review the
standard treatment of a scalar field in an FRW universe.  In section 3
we show that pair production can be obtained from the Feynman
propagator using only the late time positive-frequency modes.  The
number of particles with momentum $k$ at late times, $\langle N_k
\rangle$, turns out to be given by a simple expression:
\begin{equation}
\langle N_k \rangle = \frac{\vert \gamma_k \vert^2}{1 - \vert \gamma_k \vert^2}
\qquad {\rm where} \qquad
|\gamma_k|^2 = \left| \frac{\w_k - \wbar_k^*}{\w_k + \wbar_k} \right|^2  \;.
\end{equation}
Here $\w_k$ is a free parameter that specifies the initial state of
this momentum mode of the field.  The quantity $\wbar_k$ is computed
from out modes, $u_k(\eta)$, that are positive-frequency at late
times:
\begin{equation}
\wbar_k = i \pl_\eta \log u_k \vert_{\eta =\eta_0} \; ,
\end{equation}
where $\eta_0$ is the conformal time at which initial conditions are
specified.  In practice this technique seems more computationally
efficient than the conventional approach; in the appendix, we
demonstrate the method by calculating particle creation in various
cosmological backgrounds.  Although the approach so far is perfectly
adequate for computing pair creation, it still requires knowing the
out modes and consequently does not provide a fully first-quantized
formalism. Therefore, in section 4, we show that pair production can
be determined from the Feynman propagator in an auxiliary spacetime.
This provides a method for calculating $\wbar_k$ using only
first-quantized techniques.

\section{Scalar Field Quantization}

Consider a free minimally-coupled real scalar field $\hat{\phi}(x)$ in
a Robertson-Walker universe with metric
\[
ds^2 = a^2(\eta)\left(-d\eta^2 + h_{ij} dx^i dx^j\right)\,.
\]
Here $a(\eta)$ is the scale factor, $\eta$ is conformal time, and
$h_{ij}$ is the metric of a maximally-symmetric $n$-dimensional space.
We make the field redefinition
\begin{equation}
\hat{\phi}(x) = a^{(1-n)/2} (\eta) \hat{\chi} (x) \; ,
\end{equation}
and write $\hat{\chi}(x)$ as a sum over momentum modes:
\begin{equation}
\hat{\chi}(x) = \sum_k \lf \hat{a}_k f_k(\vec{x}) u_k (\eta) +
\hat{a}_k^\dagger f^*_k(\vec{x}) u_k^*(\eta) \rt \; .
\end{equation}
The corresponding mode functions $u_k(\eta)$ satisfy
\begin{equation}
u_k'' + V_k u_k = 0
\label{modeeqn}
\end{equation}
where
\begin{equation}
V_k(\eta) = k^2 + a^2 m^2 - \frac{1}{2} (n-1) \frac{a''}{a}
- \frac{1}{4}(n-1)(n-3) \lf \frac{a'}{a} \rt^2 \,.
\label{potentialeqn}
\end{equation}
Here $'$ indicates a derivative with respect to $\eta$.  If one thinks
of $\eta$ as position, then this is a one-dimensional Schr\"{o}dinger
equation with potential $-\frac{1}{2}V_k(\eta)$.

\subsection{Final State}
\label{FinalState}

In a general time-dependent background there is no useful sense in
which the operators $\hat{a}_k$ and $\hat{a}_k^\dagger$ annihilate and
create particles. However throughout this paper we will assume that
the universe is asymptotically adiabatic in the future.  Then the WKB
approximation becomes valid as $\eta \rightarrow \infty$, and we can
take the modes to satisfy \cite{birrelldavies}
\begin{equation}
u_k (\eta) \sim e^{-i \int^\eta d \eta' \sqrt{V(\eta')}}
\quad
\hbox{\rm as $\eta \to +\infty$.}
\label{adiabatic}
\end{equation}
This means the $u_k(\eta)$ are positive-frequency in the future, while
the $u^*_k (\eta)$ are negative-frequency, so the $\hat{a}_k$ are
operators which annihilate the preferred future, or out, vacuum:
\begin{equation}
\hat{a}_k \out = 0 \qquad \forall k \; .
\end{equation}

\subsection{Initial State}

We also need to specify the initial state of the field.  We will
consider initial states that can be characterized as follows.  First
choose a conformal time $\eta_0$ at which to fix initial conditions.
Then for each momentum $k$ choose a complex parameter $\w_k$.  The
$\w_k$ are arbitrary, aside from the fact that for reasons given below
we require ${\rm Re} \, \w_k > 0$ and $\w_k = \w_{-k}$.  Now let
$\hat{\chi}_k (\eta)$ be a Fourier component of the field,
\begin{equation}
\hat{\chi}_k(\eta) = \int dx \, f^*_k(\vec{x}) \hat{\chi}(\eta,\vec{x})
= \hat{a}_k u_k (\eta) + \hat{a}^\dagger_{-k} u^*_{-k} (\eta) \; .
\end{equation}
To specify the initial state of the field we define the operators
\begin{equation}
\hat{b}_k \equiv 
\sqrt{\frac{\w_k \, |\w_k|}{2 \, {\rm Re} \, \w_k}} \lf
\hat{\chi}_k (\eta_0) + \frac{i \hat{\pi}_k (\eta_0)}{\w_k}  \rt \; .
\label{omega_def}
\end{equation}
Here the conjugate momentum $\hat{\pi}_k (\eta) = \partial_\eta
\hat{\chi}_k$ satisfies $i[\hat{\pi}_k,\hat{\chi}_{k'}] = \delta_{k +
k'}$ as well as $\hat{\chi}_k^\dagger = \hat{\chi}_{-k}$,
$\hat{\pi}_k^\dagger = \hat{\pi}_{-k}$.  The normalization of
$\hat{b}_k$ and the condition $\w_k = \w_{-k}$ ensure that
$[\hat{b}_k, \hat{b}_{k'}^\dagger] = \delta_{kk'}$.

We take the initial state to satisfy $\hat{b}_k \in = 0$, or
equivalently
\begin{equation}
\pl_\eta \hat{\chi}_k (\eta_0) \in = i \w_k \hat{\chi}_k (\eta_0) \in \; .
\label{initialcond}
\end{equation}
The space of initial states we can consider is quite large: the values
of the parameters $\w_k$ are essentially arbitrary, so we can specify
one complex parameter per momentum mode.  By varying the $\w_k$ we can
sweep out the entire space of states that are related by Bogolubov
transformations.\footnote{Although two complex parameters $\alpha_k$
and $\beta_k$ appear in the Bogolubov transformations (\ref{bog}),
they are constrained to satisfy $|\alpha_k|^2 - |\beta_k|^2 = 1$.
Also a Bogolubov transformation which merely multiplies $\hat{b}_k$ by
a phase is redundant since it leaves $\in$ invariant.}

(Another way to characterize the initial state is to note that, as a
consequence of (\ref{initialcond}), the initial wavefunctional for the
field is Gaussian:
\begin{equation}
\Psi[\chi]\vert_{\eta = \eta_0} \sim \exp - {1 \over 2} \sum_k \omega_k
\vert\chi_k\vert^2 \; .
\end{equation}
We require ${\rm Re} \, \omega_k > 0$ so that the wavefunctional is
normalizable, and $\w_k = \w_{-k}$ so that the integrability condition
$[\hat{b}_k,\hat{b}_{k'}] \in = 0$ can be satisfied.)

\subsection{Bogolubov transformations}

We now recall some standard results \cite{birrelldavies,jacobson}.
Bogolubov transformations express the operators $\hat{a}_k$, which
annihilate the out-vacuum, in terms of $\hat{b}_k$, and vice versa:
\begin{equation}
\hat{a}_{k} = \sum_{k'} \lf \alpha_{kk'} \hat{b}_{k'} + \beta_{-kk'}^*
\hat{b}_{k'}^\dagger \rt \comma
\hat{b}_{k} = \sum_{k'} \lf \alpha^*_{kk'} \hat{a}_{k'} - \beta_{-kk'}^*
\hat{a}_{k'}^\dagger \rt \; . \label{bog}
\end{equation}
The Bogolubov coefficients $\alpha_{kk'}$ and $\beta_{kk'}$ are
determined by the overlap of the positive-frequency out-modes with the
positive or negative frequency in-modes.  The spatial modes are
orthonormal, so the Bogolubov coefficients are diagonal in $k$:
\begin{equation}
\alpha_{kk'} \equiv \alpha_k \delta_{kk'} \qquad
\beta_{kk'} \equiv \beta_k \delta_{kk'} \; .
\end{equation}
The initial state can be written as a squeezed state \cite{hkm,finn},
\begin{equation}
\label{squeezed}
\in = \prod_k C_k \exp \lf -{\frac{1}{2}} \gm_k \hat{a}_k^\dagger  
\hat{a}_{-k}^\dagger \rt \out \; ,
\end{equation}
where $C_k = (1 - |\gm_k|^2)^{1/4}$ so that $\bra {\rm in} \in = 1$.
Requiring that $\in$ be annihilated by the $\hat{b}_k$ implies (using
(\ref{bog}) and the commutation algebra) that $\gm_k = - \beta_k /
\alpha^*_k$.  Note that, by (\ref{squeezed}),
\begin{equation}
\gm_k = - {\bra {\rm out} | \hat{a}_{-k} \hat{a}_k \in \over \bra 
{\rm out} \in } \; . \label{gammapair}
\end{equation}
We see that $\gm_k$ is related to pair creation; indeed, it is the
normalized probability amplitude for the in-vacuum to evolve at late
times into a state that has two particles in it, $\hat{a}^\dagger_k
\hat{a}^\dagger_{-k} \out$. The average number of particles produced
is given by the expectation value of the number operator,
\begin{equation}
\bra {\rm in} | \hat{N}_k  \in = |
\beta_k |^2 = {| \gm_k |^2 \over 1 - | \gm_k |^2} \; , \label{number}
\end{equation}
where we have used the normalization condition $|\alpha_k|^2 -
|\beta_k|^2 = 1$.

\section{Pair Creation Using Only Out-Modes}

The Feynman propagator is defined as the vacuum expectation value of
the time-ordered product of field operators. However, the vacua
appearing on the left and right of the operator product are different:
\begin{equation}
\label{GreensEqn}
i G_F (k, \eta_1; k', \eta_2)  =  
{\bra {\rm out} | T \hat{\chi}_{k} (\eta_1) \hat{\chi}_{-k'} (\eta_2) |
{\rm in} \ket \over \bra {\rm out} \in} \; . \label{inout}
\end{equation}
Writing $\hat{\chi}_k$ in terms of modes that are positive-frequency
at late times, $\hat{\chi}_k = \hat{a}_k u_k (\eta) +
\hat{a}^\dagger_{-k} u^*_{-k} (\eta)$, we find that (for $\eta_1 >
\eta_2$)
\begin{eqnarray}
i G_F (k, \eta_1; k', \eta_2) & = &
{\bra {\rm out} |\hat{a}_{k} \hat{a}_{k'}^\dagger \in \over \bra {\rm out} \in}
u_k (\eta_1) u^*_{k'} (\eta_2) +
{\bra {\rm out} |\hat{a}_{k} \hat{a}_{-k'} \in \over \bra {\rm out} \in}
u_k (\eta_1) u_{-k'} (\eta_2) \nonumber \\
& = & 
\lf u_k (\eta_1) u^*_k (\eta_2) - \gm_k u_k (\eta_1) u_{-k} (\eta_2) 
\rt \delta_{kk'} \; . \label{GFdef}
\end{eqnarray}
To obtain the second term in the second line we used (\ref{squeezed})
and (\ref{gammapair}). Since, by (\ref{modeeqn}), $u_k(\eta) = u_{-k}
(\eta)$ we will suppress all the $k$ indices henceforth. Then in
general the Feynman propagator, expressed in terms of the out-modes,
is simply
\begin{equation}
i G_F(\eta_1, \eta_2) = u(\eta_1) u^* (\eta_2) \theta(\eta_{12}) +
u(\eta_2)u^*(\eta_1) \theta(\eta_{21})
-\gm u(\eta_1) u (\eta_2) \; . \label{iGF}
\end{equation}
The first two terms are standard but the third term is a consequence
of the in and out vacua being different. Moreover we see that the
extra term, being proportional to $\gm$, signals pair creation.

We set the boundary conditions at some time in the far past, $\eta_0$,
using (\ref{initialcond}):
\begin{equation}
\left. \pl_{\eta_2} i G_F(\eta_1, \eta_2) \right|_{\eta_2 = \eta_0}
 =  {\bra {\rm out} | \hat{\chi} (\eta_1) (\pl_{\eta} \hat{\chi})  
(\eta_0) \in \over \bra {\rm out} \in}
 =  {\bra {\rm out} | \hat{\chi} (\eta_1) (i \w) \hat{\chi}  
(\eta_0) \in \over \bra {\rm out} \in} \; .
\end{equation}
Hence the choice of initial state is encoded in the propagator as a
boundary condition at $\eta_0$:
\begin{equation}
\left. \pl_{\eta_2} i G_F(\eta_1, \eta_2) \right|_{\eta_2 = \eta_0}
= (i \w )i G_F(\eta_1, \eta_0) \quad \hbox{{\rm for} $\eta_1 > \eta_0$} \; .
\label{BCGF_eta0}
\end{equation}
Inserting (\ref{iGF}) into (\ref{BCGF_eta0}), we find that
\begin{equation}
u(\eta_1) u'^*(\eta_0) - \gm u(\eta_1) u' (\eta_0) 
= i \w \lf u(\eta_1) u^*(\eta_0) - \gm u(\eta_1) u(\eta_0) \rt \; ,
\end{equation}
so that
\begin{equation}
\gm = {\w + i \pl \ln u^*(\eta_0) \over \w + i \pl \ln u(\eta_0)}
\cdot {u^*(\eta_0) \over u(\eta_0)} \; .
\end{equation}
We can simplify this expression by defining\footnote{Throughout this
paper overbars indicate quantities associated with the reflected
problem we will introduce in section 4.  They do not indicate
complex conjugation, which we denote with a ${}^*$.}
\begin{equation}
\wbar \equiv \left . i \pl_{\eta} \ln u(\eta) \right|_{\eta = \eta_0}
\label{wbar} \; .
\end{equation}
The amount of pair creation is determined by $|\gm|^2$, which is given
by
\begin{equation}
|\gm|^2 = \left| {\frac{\w - \wbar^*}{\w + \wbar}} \right|^2
 \; . \label{gamma2}
\end{equation}
This is the equation we are after. It tells us that given $\w$ and
$\wbar$ one knows the amount of pair production. Since $\w$
parameterizes our choice of initial state, the only thing one has to
compute is $\wbar$. This one obtains from (\ref{wbar}) by
differentiating the out-modes at the point $\eta_0$ where the initial
conditions are specified.  Since one takes a logarithmic derivative,
there is no need to normalize the out-modes. More importantly, there
is no need to determine any in-modes: particle production can be
evaluated without computing Bogolubov coefficients.  We present
several examples of this economical formalism in the appendix.

\section{First-Quantized Formalism}

Our derivation so far is self-contained and sufficient for calculating
pair production.  However we do not yet have a first-quantized
formalism, since to compute $\wbar$ from (\ref{wbar}) we still need
the late-time positive-frequency modes.  In this section we show how
to obtain $\wbar$ without using mode solutions, by considering the
Feynman propagator in an auxiliary spacetime.  This will lead us to a
fully first-quantized language for discussing particle production.

One might imagine that the Green's function (\ref{GreensEqn}) could be
expressed in first-quantized language as
\begin{equation}
i G_F(\eta_1,\eta_2) = \langle \eta_1 \vert {i \over - D^2 + i \epsilon} \vert \eta_2 \rangle 
= \int_0^\infty ds \, e^{-\epsilon s} \int_{\eta(0) = \eta_2, \eta(s) = \eta_1}
\!\!\!\!\!\!\!\!\!\!\!\!\!\!\!\!\!\!\!\!\!\!\!\!\!
{\cal D}\eta(\cdot) \,\,\,\,\, e^{i S[\eta]} \;  .
\label{schwinger}
\end{equation}
Here $D^2 (\eta) = \pl^2_\eta + V(\eta)$ is the differential
(Schr\"odinger) operator that acts on the modes in (\ref{modeeqn}).
We have introduced an $i \epsilon$ prescription to make the Green's
function well-defined, and given a path integral representation
involving a Schwinger proper-time parameter $s$ and a worldline action
$S[\eta] = \int dt \left(- {1 \over 4} \dot{\eta}^2 - V(\eta)\right)$.
The basic difficulty with this representation is that, although the $i
\epsilon$ prescription picks out definite in- and out-states, in a
general curved background it is not clear exactly what those states
are (p.\ 76 in \cite{birrelldavies}).

This difficulty can be overcome with the following trick.  Consider a
separate, auxiliary problem for which we reflect the spacetime
symmetrically about $\eta_0$. We then have a new operator, $\bar{D}^2
(\eta)$, defined by
\begin{equation}
\bar{D}^2 (\eta) \equiv \left\{
	\begin{array}{ll}
	D^2 (\eta) 	        & \qquad \eta \geq \eta_0 \\
	D^2 (2 \eta_0 - \eta) 	& \qquad \eta \leq \eta_0 \; .
	\end{array} \right.
\end{equation}
The corresponding potential $\bar{V}(\eta)$ is the reflection of $V$
about $\eta_0$.  We define the Green's function in this reflected
problem with an $i \epsilon$ prescription, so that it can be computed
in first-quantized terms.
\begin{eqnarray}
&& i \bar{G}_F(\eta_1,\eta_2) = \langle \eta_1 \vert {i \over - \bar{D}^2 + i \epsilon} \vert \eta_2 \rangle 
= \int_0^\infty ds \, e^{-\epsilon s} \int_{\eta(0) = \eta_2, \eta(s) = \eta_1}
\!\!\!\!\!\!\!\!\!\!\!\!\!\!\!\!\!\!\!\!\!\!\!\!\!
{\cal D}\eta(\cdot) \,\,\,\,\, e^{i \bar{S}[\eta]} \nonumber \\
&& \bar{S}[\eta] = \int_0^s dt \lf -{1 \over 4} \lf{d \eta \over dt} \rt^2 - \bar{V} (\eta) \rt\,.
\end{eqnarray}
Notice that, as we are throwing away the part of the potential before
$\eta_0$, any early-time singularity of the potential is
irrelevant. The reflected potential may have a cusp at $\eta_0$ but
this causes no problems. The advantage of reflecting the potential is
that the spacetime is now asymptotically adiabatic in the past as well
as in the future, so the $i \epsilon$ prescription automatically
selects the preferred in- and out-vacua of the reflected problem.
That is, as $\eta_1,\eta_2 \rightarrow \pm \infty$ the Green's
function obeys the adiabatic boundary conditions considered in
sect.~\ref{FinalState}.  For example\footnote{The proof is as
follows. The Feynman propagator $i\bar{G}_F (\eta_1, \eta_2) = \langle
\eta_1 \vert {i \over -\bar{D}^2 + i \epsilon} \vert \eta_2 \rangle$.
Insert a complete set of energy eigenstates $\bar{D}^2 \vert \omega
\rangle = E(\omega) \vert \omega \rangle$ and let $\psi_\w (\eta) =
\langle \eta \vert \w \rangle$. Adiabaticity implies that for $\eta
\approx \eta_2$ we can take $\psi_\w(\eta) \approx \exp (-i \omega
\eta)$ with $E(\omega) \approx - \omega^2 + \bar{V}(\eta_2)$.  Then
\begin{equation}
i \bar{G}_F(\eta_1, \eta_2) \approx \int_{-\infty}^\infty d\omega \rho (\w) {i \over \w^2
- \bar{V}(\eta_2) + i \epsilon} \psi_\w (\eta_1) e^{i \omega \eta_2}
\end{equation}
where $\rho(\w)$ is the density of states.  The usual contour
deformation picks up the pole at $\omega = \sqrt{\bar{V}(\eta_2)} - i
\epsilon$, and (\ref{BC2}) follows.  Our approximations become exact
as $\eta_2 \to -\infty$.}
\begin{equation}
\pl_{\eta_2} \log i\bar{G}_F (\eta_1,\eta_2) \simeq i \sqrt{\bar{V}(\eta_2)}
\quad
\hbox{\rm as $\eta_2 \rightarrow -\infty$.}
\label{BC2}
\end{equation}

Suppose we are given the Feynman propagator $i \bar{G}_F$ in the
reflected problem.  It turns out that $\bar{\omega}$ can be extracted
from this auxiliary propagator.  To see this, imagine we knew the
modes $\bar{u}(\eta)$ that satisfy the reflected differential equation
$\bar{D}^2 \bar{u} (\eta) = 0$. This is just a Schr\"odinger equation
in a symmetric potential; a solution can be written in terms of the
original unreflected modes as
\begin{equation}
\bar{u}(\eta) =
\left\{ \begin{array}{ll}
      c_1 u^*(\eta) + c_2 u(\eta) & \qquad \eta \geq \eta_0 \\
      u^*(2 \eta_0 - \eta)   & \qquad \eta \leq \eta_0  \; .
\end{array} \right.
\label{ubar_def}
\end{equation}
Recall that $u$ is assumed to be positive frequency at late times, so
$\bar{u}$ is a positive frequency mode as $\eta \to - \infty$.  The
coefficients $c_1$ and $c_2$ are fixed by requiring that $\bar{u}$ and
its first derivative are continuous at $\eta_0$.  We will not need
explicit expressions for $c_1$ and $c_2$; indeed the whole point is
that we won't need to know the modes at all.

The Feynman propagator in the auxiliary problem can be expressed in
terms of the reflected modes:
\begin{eqnarray}
i\bar{G}_F (\eta_1, \eta_2) &=& 
\bar{u}(\eta_1) \bar{u}^*(\eta_2) \theta(\eta_{12}) +
\bar{u}(\eta_2) \bar{u}^*(\eta_1) \theta(\eta_{21}) 
+ \lambda_1 \bar{u}(\eta_1) \bar{u}(\eta_2) \nonumber \\
& & + \lambda_2 [ \bar{u}(\eta_1) \bar{u}^*(\eta_2) +  \bar{u}^*(\eta_1)
  \bar{u}(\eta_2)] +
\lambda_3 \bar{u}^*(\eta_1) \bar{u}^*(\eta_2)
\; , \label{gf}
\end{eqnarray}
where the constants $\lambda_i$ are coefficients of possible
homogeneous terms. Every Feynman propagator between arbitrary in- and
out-states can be written in this manner. We fix the $\lambda_i$ by
imposing boundary conditions corresponding to the initial and final
states of the reflected problem, as determined by the $i \epsilon$
prescription.  From (\ref{BC2}) it follows that $\lambda_1 = \lambda_2
= 0$. Imposing the analogous equation at late times leads to an
expression for $\lambda_3$, but the precise form is unimportant.
Having set $\lambda_1$ and $\lambda_2$ to zero, we take the derivative
of the reflected Green's function at $\eta_0$, and find
\begin{equation}
\left. \pl_{\eta_2} i\bar{G}_F(\eta_1, \eta_2) 
\right|_{\eta_1 > \eta_2 = \eta_0} 
= \lf \pl_\eta \log \bar{u}^* \rt \vert_{\eta = \eta_0} ~
i\bar{G}_F(\eta_1, \eta_0) \; .
\end{equation}
But
\begin{equation}
\partial_\eta \log \bar{u}^* \vert_{\eta = \eta_0} =
- \partial_\eta \log u \vert_{\eta = \eta_0} = i \bar{\omega}
\end{equation}
where we have used (\ref{ubar_def}) and (\ref{wbar}).  So finally we
arrive at
\begin{equation}
i \wbar = \left. \pl_{\eta_2} \ln i\bar{G}_F(\eta_1, \eta_2) 
\right|_{\eta_1 > \eta_2 = \eta_0} \; . \label{wbarGFbar}
\end{equation}
We see that we can determine $\wbar$ using only the propagator in the
auxiliary reflected problem, without making any reference to mode
functions.  Representing $i \bar{G}_F$ as a particle path integral,
this provides a fully first-quantized formalism for calculating
cosmological pair production.

\subsection{Path integrals and image charges}

We conclude by showing how to represent the original Green's function
$i G_F$ in first-quantized language.  Naively the propagator has the
path integral representation (\ref{schwinger}); the difficulty is in
understanding how to implement the correct boundary conditions at
$\eta_0$.

To overcome this difficulty we express the original Green's function
using the method of images, as
\begin{equation}
i G_F(\eta_1, \eta_2) = i \bar{G}_F(\eta_1, \eta_2) + q i
\bar{G}_F(\eta_1, 2 \eta_0 - \eta_2) \; , \label{image_charge}
\end{equation}
where $q$ is an image charge that we will determine shortly.  For
$\eta_1, \eta_2 > \eta_0$ note that $i G_F$ is indeed a Green's
function for the operator $D^2(\eta)$. It remains to implement the
boundary condition at $\eta_0$ by choosing $q$ appropriately.  From
(\ref{BCGF_eta0}) we require
\begin{equation}
\left. \pl_{\eta_2} i G_F(\eta_1, \eta_2) \right|_{\eta_1 > \eta_2 = \eta_0}
= (i \w )i G_F(\eta_1, \eta_0)
\end{equation}
while from (\ref{image_charge}) we have
\begin{equation}
i G_F(\eta_1, \eta_0) = (1+ q) i \bar{G}_F(\eta_1, \eta_0)
\label{plus_q}
\end{equation}
and
\begin{equation}
\pl_{\eta_2} i \left. G_F(\eta_1, \eta_2) \right|_{\eta_1 > \eta_2 =
\eta_0} = (1- q) \pl_{\eta_2} i \left. 	\bar{G}_F(\eta_1,
\eta_2) \right|_{\eta_1 > \eta_2 = \eta_0} \; .
\label{minus_q}
\end{equation}
Using (\ref{wbarGFbar}), we find that the image charge is given by
\begin{equation}
q = {\wbar - \w \over \wbar + \w} \; .
\end{equation}

The method of images also provides a way of representing the original
propagator in terms of a particle path integral.  Use
(\ref{image_charge}) to represent $i G_F$ as a sum of two particle
path integrals in the auxiliary reflected potential.  In each path
integral fold the particle paths across $\eta = \eta_0$, to obtain
particle worldlines that are restricted to satisfy $\eta(s) \geq
\eta_0$.  Note that this folding leaves the worldline action
invariant.  By adding the two path integrals one can represent $i G_F$
as a single path integral, just as in (\ref{schwinger}), but where the
particle paths are restricted to satisfy $\eta(s) \geq \eta_0$, and
where the boundary conditions at $\eta = \eta_0$ are enforced by
weighting paths according to the rule
\begin{eqnarray}
\nonumber
&& e^{i S[\eta]} \qquad\,\,\, \hbox{\rm for paths that touch $\eta_0$ an even number of times} \\
\label{BCrule}
&& q e^{i S[\eta]} \qquad \hbox{\rm for paths that touch $\eta_0$ an odd number of times}
\end{eqnarray}
Similar constructions to enforce boundary conditions on the path
integral are discussed in \cite{PPIbc1,PPIbc2}.

\section{Conclusions}

To summarize: given a cosmological background and a set of initial
conditions specified by the parameters $\omega_k$, our recipe is to
first compute the Green's function in the reflected potential $i
\bar{G}_F$.  This determines the quantities $\wbar_k$ and hence the
ratio of Bogolubov coefficients $|\gm_k|$.  This is all that is needed
for particle production, since $\vert \gamma_k \vert$ controls the
amount of particle production.  Since $i \bar{G}_F$ can be computed
from a particle path integral, we have achieved our goal of finding a
first-quantized formalism for computing cosmological particle
production.

Our approach has nice conceptual advantages over conventional
Bogolubov techniques.  The formation of a pair of particles from the
vacuum intuitively suggests a U-shaped particle worldline, with the
two endpoints of the U marking the two particle positions at late
times, and with the bottom of the U marking (heuristically) the time
of pair creation.  Our approach makes this intuition precise, by
giving a prescription (\ref{BCrule}) for obtaining the appropriate
Feynman propagator $i G_F$ from an integral over particle paths.
Moreover, in our approach the generalization to string theory is
immediate: replace ``particle worldlines'' with ``string
worldsheets''.

\bigskip
\noindent
\begin{center}
{\bf Acknowledgements}
\end{center}

\noindent 
We are very grateful to Nori Iizuka for collaboration in the early
stages of this project. We thank Brian Greene, Gilad Lifschytz, and
Govindan Rajesh for valuable discussions. {DK} and {MP} are grateful
to the organizers of the Amsterdam and Aspen workshops where part of
this work was completed, and {DK} is grateful to the Rutgers theory
group for its hospitality. {AH} is supported by DOE grant
DE-FG02-92ER40699, {DK} is supported by DOE grant DE-FG02-92ER40699
and by US--Israel Bi-national Science Foundation grant \#2000359, and
{MP} is supported in part by DOE grant DF-FCO2-94ER40818.

\section{Appendix: Examples}

In this appendix, we demonstrate how the formalism of section 3 is
applied by calculating pair production in various cosmological
backgrounds. In practice, calculating particle production boils down
to determining the late-time positive-frequency modes and taking their
derivative at the point $\eta_0$ at which the initial conditions are
imposed.

\subsection{The Milne universe}

The line element for a two-dimensional Milne universe is
\begin{equation}
ds^2 = -dt^2 + s^2 t^2 dx^2 = e^{2s \eta} \lf - d\eta^2 + dx^2 \rt \; ,
\end{equation}
where $-\infty < \eta,x < \infty$. This space is simply the upper
quadrant of Minkowski space. Consider a massive scalar field
propagating in this geometry. The wave equation is
\begin{equation}
\lf \pl^2_\eta + k^2 + m^2 e^{2s \eta} \rt u_k(\eta) = 0 \; .
\end{equation}
At early times the solutions are just plane waves so the natural
choice for the in-vacuum is to set $\w = k$.  We will adopt this
choice, which corresponds to the conformal vacuum
\cite{birrelldavies}, although we could easily consider different
initial conditions.  At late times the positive-frequency modes are
Hankel functions:
\begin{equation}
u_k(\eta) = N H^{(2)}_{\nu} \lf (m/s) e^{s\eta} \rt \; ,
\end{equation}
where $\nu = ik/s$ and $N$ is a normalization constant that we will
not need to determine. Suppose we specify initial conditions in the
infinite far past, $\eta_0 \to - \infty$. Then we need the behavior of
the out-mode as $\eta \to -\infty$.  The Hankel function is
proportional to $J_{-\nu}(x) - e^{-\pi k/s} J_{\nu} (x)$ and, as $x
\to 0$,
\begin{equation}
J_{\nu} (x) \approx (x/2)^\nu / \Gm(1+\nu) = A e^{ik\eta} \; ,
\end{equation}
where $A$ is an $\eta$-independent constant. Particle production is
controlled by
\begin{equation}
\wbar = i \pl_\eta \ln H^{(2)}_\nu (0)
  = k {A^* e^{-i k \eta} + A e^{-\pi k/s} e^{ik \eta}
  \over A^* e^{-ik \eta} - A e^{-\pi k/s} e^{ik \eta}} \; .
\end{equation}
{}From this we obtain $\gm_k$:
\begin{equation}
|\gm_k| = \left | {k - \wbar^* \over k + \wbar} \right | = e^{-\pi
  k/s} \; .
\end{equation}
The occupation number of a mode with momentum $k$ is, using
(\ref{number}),
\begin{equation}
\bra {\rm in} | \hat{N}_k \in  = {1 \over e^{2 \pi k/s} - 1} \; .
\end{equation}
This indicates that modes in the Milne universe are, up to mass
corrections, populated according to a Planck spectrum with temperature
$T = s/2 \pi$.  This is in accord with the standard result that the
conformal vacuum appears thermal to a comoving observer
\cite{birrelldavies,bunch,bcf}.

\subsection{A doubly-asymptotically-static universe}

Consider a two-dimensional universe with scale factor
\begin{equation}
a^2(\eta) = A + B \tanh \rho \eta \; ,
\end{equation}
where $A, B, \rho$ are positive constants. The wave equation is
\begin{equation}
\lf \pl^2_\eta + k^2 + m^2 \lf A + B \tanh \rho \eta \rt \rt u_k(\eta)
= 0 \; .
\end{equation}
This cosmology has no big bang; it is asymptotically static in the
distant future as well as in the distant past. As $\eta \to \pm
\infty$ the mode solutions behave like plane waves, albeit with
different frequencies:
\begin{equation}
\w_{\rm out} = (k^2 + m^2 (A+B))^{1/2} \qquad \w_{\rm in} = (k^2 + m^2
(A-B))^{1/2} \; .
\end{equation}
It is useful to define $\w_{\pm} = (\w_{\rm out} \pm \w_{\rm in})/2$. 
The mode function with positive frequency at late times is
\begin{equation}
u_k(\eta) = N e^{-i \w_+ \eta} (\cosh \rho \eta)^{-i \w_-/\rho}
F(a,b;c;(1/2) (1- \tanh \rho \eta)) \; ,
\end{equation}
where $F$ is a hypergeometric function with
\begin{equation}
a = 1 + i\w_-/\rho \qquad b = i \w_-/\rho \qquad c = 1 +
i\w_{\rm out}/\rho \; .
\end{equation}
In the conventional approach \cite{bernardduncan} one would have to
determine the modes that are positive-frequency at early times; these
are just as messy as the late-time modes.  One then has to find the
linear transformation that relates the two sets of modes. But in our
approach it suffices to note that the positive-frequency modes behave
as plane waves at early times.  Thus a natural choice is to set $\w =
\w_{\rm in}$.  To compute the particle production we need the
logarithmic derivative of the out-modes evaluated in the far
past. Defining $x = (1/2) (1- \tanh \rho \eta)$, we expand $F$ around
$x = 1$:
\begin{equation}
F(a,b;c;x) \approx (1-x)^{c-a-b} A + ... + B + ... \; ,
\end{equation}
where the dots indicate terms that are unimportant as $x \to 1$. Here
A and B are
\begin{equation}
A = {\Gm(c-a-b) \Gm(c) \over \Gm(a) \Gm(b)} \qquad B = {\Gm(c-a-b)
\Gm(c) \over \Gm(c-a) \Gm(c-b)} \; .
\end{equation}
Then as $\eta \to -\infty$ or $x \to 1$, we have
\begin{equation}
\wbar = \w_{\rm in} \left. \lf 1 - {2 A (1-x)^{i w_{\rm in}
  /\rho} \over (1-x)^{i w_{\rm in}/\rho} A + B } \rt \right|_{x=1} \; . 
\end{equation}
Some simple algebra yields
\begin{equation}
|\gm| = |A/B| = \left | {\Gm(i\w_+ /\rho) \Gm(1+ i\w_+/\rho)
  \over \Gm(i\w_- /\rho) \Gm(1+ i\w_-/\rho)} \right | \; ,
\end{equation}
which agrees precisely with the result obtained by conventional methods
\cite{bernardduncan,birrelldavies}.

\subsection{Power-law FRW}

In the two preceding examples the mode solutions reduce to plane
waves at early times. A natural choice for the in-vacuum was therefore
one in which the annihilation operator $\hat{b}_k$ is paired with
$\exp(-i \w_{\rm in} \eta)$ so that $\w = \w_{\rm in}$. In a general
cosmological background, however, there is no preferred vacuum state
at early times, and $\w$ is just a parameter that characterizes the
initial state. For example consider a 3+1-dimensional Robertson-Walker
spacetime with a power-law scale factor, $a(t) \sim t^c$. We consider
$0 < c < 1$ so the expansion is decelerating. The line element reads
\begin{equation}
ds^2 = -dt^2 + a^2(t) d\Sigma_3^2 = C^2 \eta^{2c \over 1-c} (-d\eta^2 + d
\Sigma_3^2) \; ,
\end{equation}
where $0 < \eta < \infty$ and $C$ is an unimportant constant. A
massless minimally-coupled real scalar field $\phi(x)$ can be
conveniently written as $\phi(x) = \chi(x)/a(\eta)$.  The mode
equation is
\begin{equation}
\lf \pl^2_\eta + k^2 - {\nu^2 - 1/4 \over \eta^2} \rt \chi_k(\eta) = 0
\comma \nu = {3c-1 \over 2(1-c)} \; .
\end{equation}
If $\nu^2 \neq 1/4$ the solutions do not resemble plane waves at
early times and there is no natural choice for $\w$. Let us pick some
arbitrary time $\eta_0$ at which to fix our choice of initial state.
Define $\hat{\chi}_k (\eta_0) = \hat{\chi}_0$ and $(\pl_\eta
\hat{\chi}_k)(\eta_0) = \hat{\pi}_0$. As before, define an operator
\begin{equation}
\hat{b}_k = N_k \lf \hat{\chi}_0 + i {\hat{\pi}_0 \over \w_k} \rt \; ,
\end{equation}
and set $\hat{b}_k \in = 0$. This implies that
\begin{equation}
\hat{\pi}_0 \in = i \w_k \hat{\chi}_0 \in \; .
\end{equation}
The out-vacuum is unique, since the field evolves adiabatically at
late times.  The positive-frequency modes at late times are
\begin{equation}
u_k(\eta) = N \sqrt{\eta} H^{(2)}_\nu (k \eta) \; ,
\end{equation}
so that
\begin{equation}
\wbar = i \left. \pl_\eta \ln \sqrt{\eta} H^{(2)}_\nu (k \eta)
\right |_{\eta = \eta_0} \; .
\end{equation}
To see what happens as $\eta_0 \to 0$, expand $u_k(\eta)$ around
$\eta=0$, keeping the lowest order real and imaginary terms:
\begin{eqnarray}
\wbar  &\approx& i\pl_{\eta} \ln \left. \left[ A \lf {k \eta \over 2} \rt^{\nu 
+ 1/2} + iB \lf {k \eta \over 2} \rt^{1/2 - \nu} \right] \right
|_{\eta = \eta_0} \nonumber \\
&=& i\pl_{\eta} \ln \left. \left[ \lf 1 - i {A \over B} \lf {k 
\eta \over 2} \rt^{2\nu} \rt
\lf iB \lf {k \eta \over 2} \rt^{1/2 - \nu} \rt \right] \right |_{\eta
  = \eta_0} \nonumber \\
&\approx&  {k \nu \Gm (1 - \nu) \over \Gm (1+\nu)} \sin (\nu \pi) \lf 
{k \eta_0 \over 2} \rt^{2 \nu -1} + {2 \nu - 1 \over 2 i \eta_0} \; ,
\end{eqnarray}
where $A = \sqrt{2 \over k} {1 \over \Gm (1+\nu)}$, $B = \sqrt{2 \over
k} {\csc (\nu \pi) \over \Gm (1-\nu)}$, and in going from the second
to the third line we expanded the log about $1$.  Particle production
is controlled by
\begin{equation}
|\gm| \approx  \left| {\w - {k \nu \Gm (1 - \nu) \over \Gm (1+\nu)} \sin (\nu 
\pi) \lf {k \eta_0 \over 2} \rt^{2 \nu -1} +  i{({1 \over 2} 
- \nu)\over \eta_0} \over \w + {k \nu \Gm (1 - \nu) \over \Gm (1+\nu)}
\sin (\nu \pi) \lf {k \eta_0 \over 2} \rt^{2 \nu -1} +  i{({1 \over 2}
- \nu) \over \eta_0}} \right|  \; .
\end{equation}
This approaches unity for $\nu^2 \neq 1/4$, since the imaginary term
dominates as $\eta_0 \to 0$. Hence the pair production diverges, a
result which perhaps is merely a sign that the initial condition
should not be applied strictly at $\eta_0 = 0$ where the geometry is
singular. However, for $\nu = \pm 1/2$, which corresponds to $c=0$ (a
static universe) or $c=1/2$ (a radiation-dominated universe), $|\gm|$
simplifies greatly to
\begin{equation}
|\gm| = \left\vert{k - \w \over k + \w}\right\vert.
\end{equation}
This result follows from the fact that for $\nu^2 = 1/4$ the Hankel
function reduces to a plane wave with energy $k$. If one sets $\w = k$
then $\in$ is identical to $\out$, so there should be no particle
production, and indeed for this value of $\w$ we see that $\gm$
vanishes.

\end{document}